\documentclass{elsart}

\usepackage{graphicx}              

\begin{document}

\begin{frontmatter}

\title{Enhanced Production of Neutron-Rich Rare Isotopes in the Reaction of 25 MeV/nucleon 
       $^{86}$Kr  on  $^{64}$Ni }

\author[]{G.A. Souliotis}$^{}$, \thanks{\footnotesize  
             E-mail address: soulioti@comp.tamu.edu (G.A. Souliotis).}
\author[]{M. Veselsky},              
\author[]{G. Chubarian},     
\author[]{L. Trache},
\author[]{A. Keksis},
\author[]{E. Martin},
\author[]{A. Ruangma},
\author[]{E. Winchester}, and
\author[]{S. J. Yennello}.

\address{Cyclotron Institute,
           Texas A\&M University, College Station, TX 77843 }

\begin{abstract}

The cross sections and velocity distributions of projectile-like 
fragments  from the reaction of 25 MeV/nucleon $^{86}$Kr + $^{64}$Ni 
have been measured using the MARS recoil separator at Texas A\&M,  with 
special emphasis  on the neutron rich isotopes. Proton-removal and 
neutron pick-up isotopes have been observed with large cross sections. 
A model of deep-inelastic transfer (DIT) for the primary interaction stage  
and the  statistical evaporation code GEMINI for the deexcitation stage
have been used to describe the properties of the product  distributions.
The results have also been  compared with the 
EPAX parametrization of high-energy fragmentation yields. 
The experimental data show an enhancement  in the production of  neutron-rich 
isotopes close to the projectile, relative to the predictions of DIT/GEMINI 
and the expectations of EPAX. We  attribute this enhancement mainly to the 
effect of the extended neutron distribution (neutron ``skin'') of  the $^{64}$Ni
target  in  peripheral interactions of  $^{86}$Kr with $^{64}$Ni.
The large cross sections of such  reactions near  the Fermi energy, 
involving  peripheral nucleon exchange, suggest that, not only the N/Z of 
the projectile and the target, but also the N/Z distribution at the nuclear 
surface  may  properly be  exploited in the production of  neutron-rich rare isotopes.
This synthesis approach may  offer  a fruitful pathway to  extremely 
neutron-rich  nuclei, towards  the neutron-drip line.

\end{abstract}


\end{frontmatter}

\section{Introduction}

Synthesis and investigation  of very neutron-rich nuclides are of 
exceptional 
importance   to advance  our understanding  of  nuclear structure 
and properties at the extreme isospin limit of the nuclear landscape.
Many of these nuclides play a key role in  stellar 
nucleosynthesis, especially in  the r-process.
The production and separation of these nuclides represent an essential 
ingredient  in regards to current or future radioactive beam facilities
(see e.g. \cite{RIA,EURISOL}).

Fission or spallation are prolific ways to generate
a variety of neutron-rich nuclides (see e.g. \cite{Geissel,Morrissey}).
In addition, projectile fragmentation of n-rich beams \cite{Morrissey}
at high or intermediate energy (well above the Fermi energy)
has been efficiently employed to produce  radioactive beams 
of n-rich nuclei. The so-called ``cold'' projectile fragmentation has been
described by abrasion/ablation models which provide a phenomenological
basis  for extrapolation to even more exotic n-rich species \cite{Friedman}.
Based on available fragmentation data, a detailed parametrization of the 
fragmentation cross sections from high-energy reactions has been developed
\cite{EPAX} and is commonly used for planning of radioactive beam experiments.
In high or medium energy projectile fragmentation reactions, the production of 
n-rich rare isotopes is based on  a more or less ``clean-cut'' removal
of protons from the projectile.  In such reactions, the target has almost no
effect on the production cross sections, apart from a geometric factor.
One-neutron pick-up products are produced with very small cross sections
\cite{Weber}.

In contrast to high-energy reactions, 
the effect of the target neutron-to-proton ratio (N/Z)  in n-rich rare isotope
production has been shown  in  multinucleon transfer reactions 
close to the Coulomb barrier. For example, large cross sections for 
several neutron pick-up channels, along with proton stripping channels
have been observed in  the reaction of $^{64}$Ni projectiles with 
$^{238}$U above the Coulomb barrier \cite{Corradi}.
The effect of the projectile and target N/Z in the 
production of projectile-like  fragments from $^{32,34}$S (E/A=6--20 MeV/nucleon)
on  $^{12}$C and $^{197}$Au has recently been reported in \cite{Tarasov},
following earlier work in this energy range \cite{Artukh,Volkov}.
Deep inelastic collisions around the Coulomb barrier 
have also been used to produce and study n-rich rare earth isotopes
at high spin states \cite{IYLee}. 
Around the Fermi energy (20--40 MeV/nucleon) a number of n-rich yield
measurements   of heavy projectiles (e.g. $^{40}$Ar, $^{86}$Kr) exist in the literature 
\cite{Bacri,Bazin}, but in these measurements no production cross sections have 
been reported. Cross sections have been reported for the reaction 70 MeV/nucleon $^{86}$Kr 
+ $^{27}$Al \cite{Pfaff}, but the n-rich side of the fragment distribution was 
not covered.  

In the present work, we have performed a systematic high-resolution spectrometric
study of the production cross sections and the velocity distributions 
of projectile-like  fragments from the reaction  25 MeV/nucleon 
$^{86}$Kr + $^{64}$Ni with particular attention on the n-rich nuclides. 
In this study,  we observed  enhanced 
production cross sections of n-rich fragments near the projectile
and the formation  of several neutron pick-up products, along with proton 
stripping products. 
The paper is organized as follows: Section 2 describes the experimental setup,
the measurements and the data analysis procedures. In Section 3, the experimental
results are presented and compared with model predictions.
Finally, in Section 4, a summary and conclusions are given.

\section{Experimental Method and Data Analysis}

The present study was performed at the Cyclotron Institute of Texas A\&M
University. A  25 MeV/nucleon  $^{86}$Kr$^{22+}$ beam 
from the K500 superconducting cyclotron, with a typical current of $\sim$5 pnA, 
interacted with an isotopically enriched (98\%) $^{64}$Ni target of  thickness 4 mg/cm$^{2}$.
The reaction products were analyzed with  the MARS recoil separator \cite{MARS}.
The primary beam struck the target at 0$^{o}$ relative to the optical 
axis of the spectrometer.  The direct beam was collected in  a small square 
Faraday cup approx. 30 cm after the target,  blocking the angular range 0.0--1.0$^{o}$.
The fragments were accepted in the remaining angular opening of MARS:  1.0--2.7$^{o}$ 
(the angular acceptance of MARS is 9 msr \cite{MARS}). 
This angular range  lies inside the grazing angle of 3.6$^{o}$  \cite{Wilcke} 
for the present reaction. 
An  Al foil (1 mg/cm$^2$) was positioned after the Faraday cup to reset 
to equilibrium the ionic charge states of the reaction procucts.
MARS optics \cite{MARS} provides one  intermediate dispersive  image and a 
final achromatic image (focal plane) and offers a  momentum acceptance of  
4\%.  

At the focal plane, 
the fragments were collected in a large area (5$\times$5 cm) three-element 
($\Delta $E$_{1}$,  $\Delta $E$_{2}$, E) Si detector telescope. 
The $\Delta$E$_{1}$ detector was a position-sensitive Si strip detector 
of 63 $\mu$m 
thickness whereas the $\Delta$E$_{2}$ and the E detector were  
single-element Si detectors of
150 and  950 $\mu$m, respectively.
The position information from the $\Delta $E$_{1}$ strips provided a continuous 
monitoring of the focusing and collection of the fragments at the various 
settings of the separator.
Time of flight was measured between two PPACs (parallel plate avalanche 
counters) \cite{Greg}
positioned at the dispersive image and at the focal plane, respectively, 
and separated by a distance of 13.2  m. 
The PPAC at the dispersive image was also  X--Y  position sensitive  and  
used  to record 
the position of the reaction products. The horizontal position, along with NMR
measurements of the field of the MARS first dipole, 
was used to determine the magnetic rigidity $B\rho $ of the particles. 
Thus, the reaction products were characterized by an event-by-event measurement
of dE/dx, E, time of flight, and magnetic rigidity. 
The response of the spectrometer/detector system 
to ions of known atomic number Z, mass number A, ionic charge q and 
velocity was calibrated using low intensity primary beams of 
$^{40}$Ar, $^{44}$Ca  and  $^{86}$Kr at 25 MeV/nucleon. 
To cover the N/Z and velocity range of the fragments, a series of measurements
was performed at overlapping magnetic rigidity
settings in the range 1.6--2.0 Tesla-meters.

\begin{figure}[p]                    

\includegraphics[width=8.0cm,height=8.0cm]{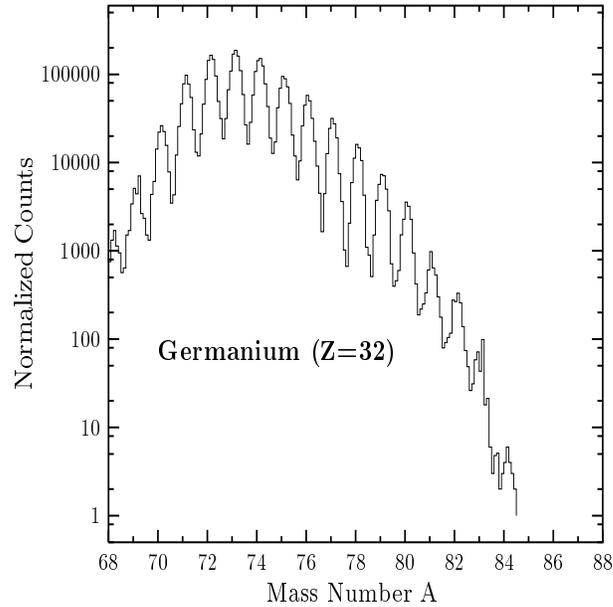}
\centering
\vspace{0.0cm}

\caption{ Mass histogram of Germanium (Z=32) isotopes }

\label{mass}
\end{figure}

The determination of the atomic number Z was based on the energy loss of the 
particles in the first $\Delta E$ detector \cite{Hubert} and their velocity,
with a resulting resolution (FWHM) of 0.5 Z units for near-projectile
fragments.
The ionic charge $q$ of the particles after the Al stripper, was obtained from
the total energy E$_{tot}$=$\Delta$E$_1$+$\Delta$E$_2$+E, the velocity and 
the magnetic rigidity
according to the expression: 
\begin{equation}
q=\frac{3.107}{931.5}\frac{E_{tot}}{B\rho (\gamma -1)}\beta \gamma
\label{q_eqn}
\end{equation}
where E$_{tot}$ is in MeV, B$\rho $ in Tm, $\beta =\upsilon /c$ and $\gamma
=1/(1-\beta ^2)^{\frac 12}$. 
The measurement of the ionic charge q had a resolution of 0.4 units (FWHM).
Since the ionic charge must be an integer, we assigned integer
values of q for each event by putting windows ($\Delta q=0.4$) 
on each peak of the q spectrum. 
Using the magnetic rigidity and velocity measurement, the mass-to-charge 
A/q ratio  of each ion was obtained from the expression: 
\begin{equation}
A/q = \frac{B\rho }{3.107\beta \gamma }  \label{Aq_eqn}
\end{equation}
Combining the q determination with the A/q measurement, the mass A
was obtained as:
\begin{equation}
A = q_{int} \times A/q  \label{A_eqn}
\end{equation}
(q$_{int}$ is the integer ionic charge determined as above) with an 
overall resolution  (FWHM) of about 0.6 A units (Fig. \ref{mass}).


Combination and appropriate normalization of the data at the various magnetic
rigidity settings of the spectrometer provided fragment  distributions with respect to 
Z, A, q and velocity. Correction of missing yields caused by charge changing 
at the  PPAC (positioned at the dispersive image) was performed
based on the equilibrium charge state prescriptions of Leon et. al. \cite{Leon}. 
(The overall data reduction procedure was similar to that followed in earlier 
work on $^{197}$Au fragmentation and was described in detail in \cite{AuPLF}.)
The distributions were subsequently summed over all values of q. 
It should be pointed out  that the resulting distributions in Z, A and velocity 
are the fragment yield distributions in the reaction  angle interval 1.0--2.7$^{o}$ 
in the magnetic rigidity range 1.6--2.0 T\,m.
Fig. \ref{mass} shows, as an example, the  mass spectrum of Germanium (Z=32) isotopes 
in full resolution.

\section{Results and Discussion}

\begin{figure}[p]                    

\includegraphics[width=10.1cm,height=15.5cm]{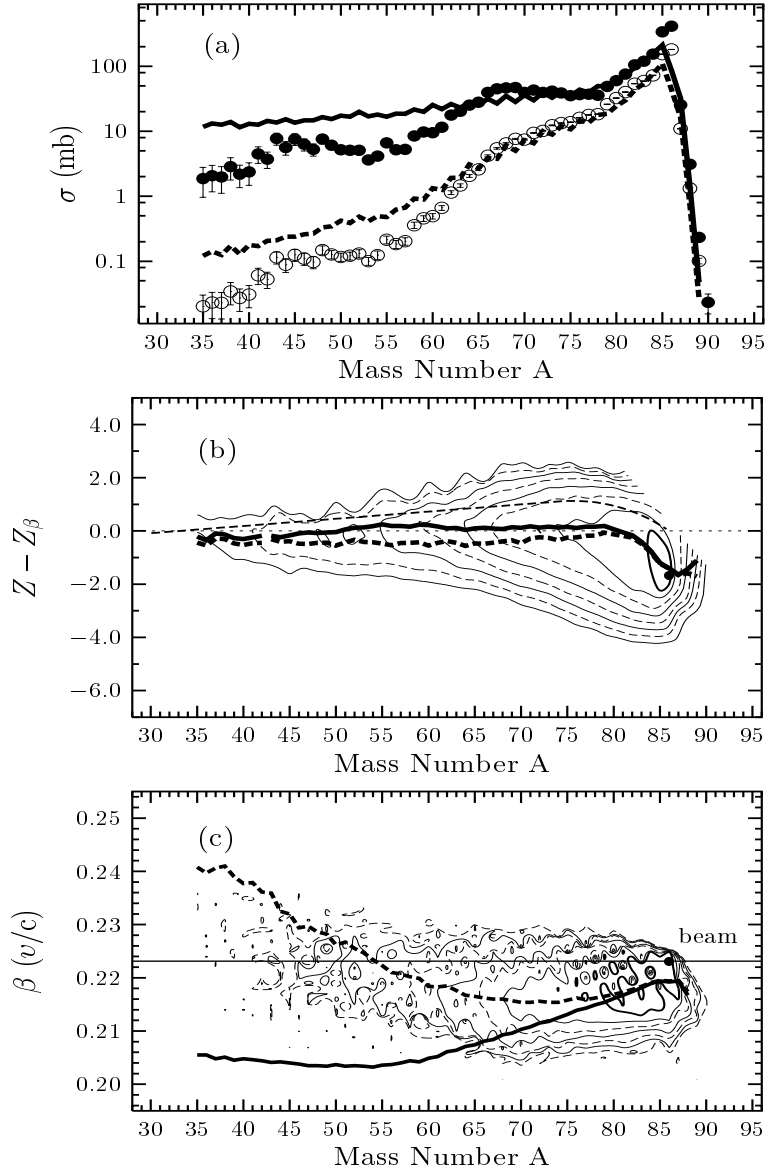}
\centering
\vspace{0.0cm}

\caption{ 
    Fragment distributions for the reaction  
    25 MeV/nucleon  $^{86}$Kr + $^{64}$Ni.
(a) - isobaric yield distribution. The data are shown as solid circles (total
    cross sections) and open circles (with acceptance cut). 
    The full line is the result of DIT/GEMINI (see text). 
    The dashed  line is  the result of the  same calculation as the full line, 
    but with a cut corresponding  to the angular and momentum acceptance 
    of the spectrometer.  
(b) - yield distributions 
    as a function of Z (relative to the line of $\protect\beta$ stability,
    Z$_{\protect\beta}$) and A.
    Highest yield contours are plotted with thicker lines. Successive contours
    correspond to a decrease  of the yield by a factor of 2.
    The calculated values from DIT/GEMINI are shown as i) thick full line:
    without acceptance cut and, ii) thick dashed line: with acceptance cut.
    Thin dashed line: EPAX parametrization. 
(c) - velocity vs. mass distributions.
    Data are shown as contours as in (b). The thick lines are as in (b).
    The  horizontal full line represents the  beam velocity.
   }

\label{gross}

\end{figure}

\begin{figure}[p]                           

\includegraphics[width=10.1cm,height=15.5cm]{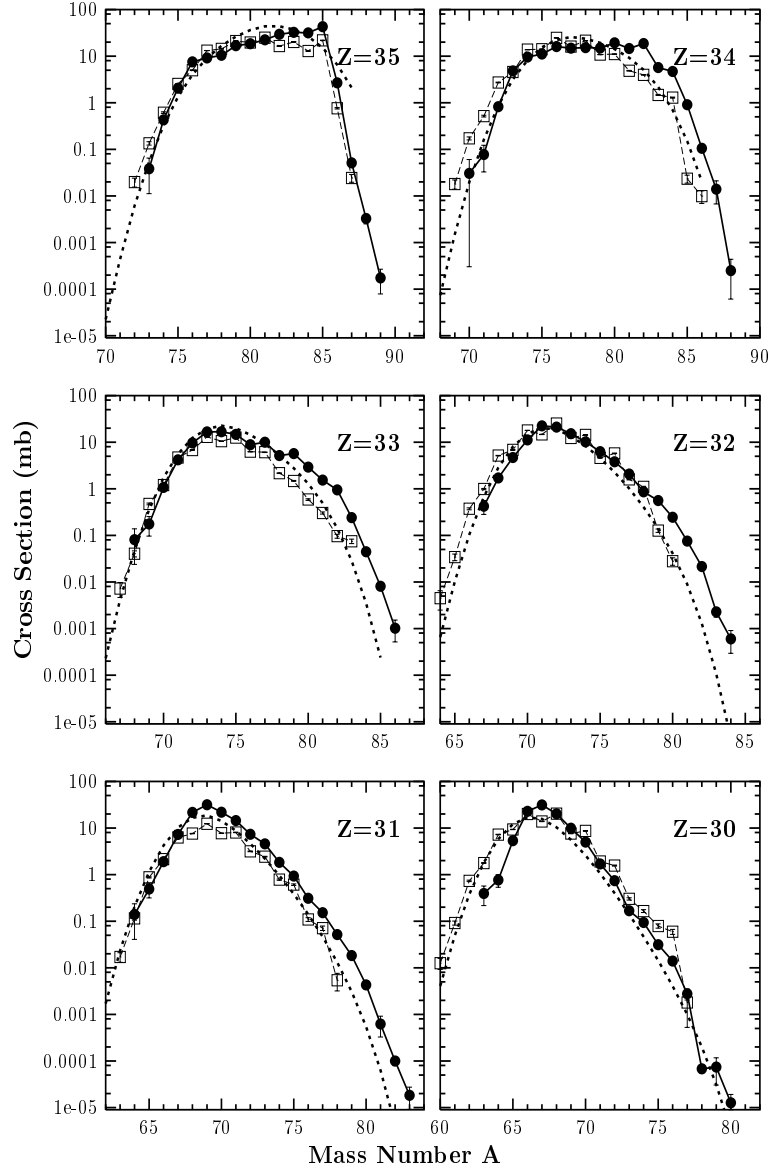}
\centering
\vspace{0.0cm}

\caption{  Mass distributions of several elements from the reaction of 
           25 MeV/nucleon $^{86}$Kr with $^{64}$Ni.
	   The present  data
           are shown by full circles. Open squares are simulations according 
	   to  DIT/GEMINI and the dotted  line is from the high-energy 
	   parametrization EPAX
           (see text). }
\label{z1}
\end{figure}

The gross features of the distributions of projectile fragments from the 
present reaction are described in Fig. \ref{gross}.
The detailed mass distributions of elements Z=35 to Z=30 are presented 
in Fig. \ref{z1}.  Before further discussion of the data, we will give
an outline of the calculations we performed for this reaction using
a phenomenological model appropriate for this energy regime.
 
The primary interaction stage was modeled with  the  deep inelastic 
transfer code  of Tassan-Got and Stephan \cite{DIT} in which stochastic
nucleon exchange was  assumed for  the orbital angular momentum range 
$\ell$=100--520. 
This DIT model has been successfully applied to describe the primary interaction
stage in studies of projectile multifragmentation of $^{28}$Si on $^{112,124}$Sn
around the Fermi energy \cite{MV0}.
Events corresponding to trajectories in which the 
projectile--target overlap exceeded 3 fm   were rejected.
Following the creation of the primary fragments by the DIT mechanism,
the statistical de-excitation of the excited primary fragments was simulated
using GEMINI \cite{GEMINI}.
This statistical deexcitation code uses Monte Carlo techniques
and the Hauser-Feshbach formalism to calculate the probabilities for 
fragment emission with Z$\leq$2. Heavier fragment emission 
probabilities are  calculated using the transition state formalism of 
Moretto \cite{Moretto}.
In the GEMINI calculations, we used Lestone's temperature dependent 
level density parameter \cite{Lestone}, a fading of shell corrections with
excitation energy and we enabled IMF emission.
Each partial-wave distribution was appropriately weighted  
and combined to give the overall fragment A, Z  and velocity distributions.
The results of the DIT/GEMINI  calculation were also  filtered by  the angular 
and B$\rho$ acceptance of the spectrometer.
The predictions of this calculation are compared with the present data in the 
following paragraphs.

 
In Fig. \ref{gross}a the mass yield curve is presented.
The measured data, normalized for beam  current and target thickness 
are given in mb and presented as open circles.
The result of the DIT/GEMINI calculation, filtered by the spectrometer
acceptance is given by the dashed line, whereas the full line 
gives the total (unfiltered) yield. 
A comparison of the measured yields (open symbols) to the calculated filtered yields
(dashed line) shows excellent agreement for the heavier fragments (A$>$65).
The  discrepancies for lower mass fragments are mainly due to incomplete coverage of the
measured data for this mass range.
Using the ratio of filtered to unfiltered calculated yield for each mass,
correction factors (whose magnitude is inferred from Fig. \ref{gross}a) 
for the acceptance of the spectrometer
were obtained as a function of mass and were applied to the measured yield data 
to obtain the total yield, given by the full circles in Fig. \ref{gross}a.
The systematic uncertainty in the extraction of  absolute cross sections 
by this procedure is estimated to be about  40\% (FWHM).
The correction factors were also  employed to obtain total isotope production 
cross sections (Fig. \ref{z1}) from the measured  yields. 
A comparison of the extracted total isobaric cross sections  (full circles) 
to the calculated unfiltered  yields (full line) again shows  fair agreement, 
except for the lower mass fragments (A$<$65).

In Fig. \ref{gross}b,  the measured yield distributions 
as a function of Z (relative to the line of $\protect\beta$ stability,
Z$_{\protect\beta}$) and A are presented as contour lines.
The line of stability is calculated as: Z$_{\beta }$ = A/(1.98 + 0.0155A$^{2/3}$)
\cite{Marmier}.
The calculated values from DIT/GEMINI are shown as  a thick full line
(without acceptance cut) and  as a thick dashed line (with spectrometer 
acceptance cut).
The thin dashed line is from the  EPAX parametrization \cite{EPAX} of 
relativistic fragmentation data and is given here for comparison.
From this Figure we see that very neutron-rich nuclides (up to 4 Z units
away from stability for isobars near the projectile) are produced.
The DIT/GEMINI calculation describes reasonably well the experimental data.
Finally, 
the locus of the relativistic fragmentation data is towards more 
proton-rich fragments than  the data in this energy regime.

Finally, in Fig. \ref{gross}c, the velocity vs. mass distributions 
are given. The data  are again shown as contours. 
The thick full line is from the DIT/GEMINI calculation without acceptance cut and 
the dashed line is with  acceptance cut. 
This mass resolved velocity distribution exhibits correlations  characteristic of 
quasi-elastic interactions (for near projectile fragments)  and deep inelastic 
interactions (for fragments further away from the projectile) as revealed by the 
continuous fragment ridge with monotonically decreasing velocities. For lower masses
(A$<$65) the fragment velocities seem to increase with decreasing mass. This is due to the 
fact that the spectrometer settings were such that only  forward moving  fragments 
coming from decay of highly excited primary  products were observed.   
In addition, comparing the filtered DIT/GEMINI calculations with the data, we see that the  
calculation is  able to describe well  the observed velocity distributions in the whole 
mass range.
In summary, based on the comparisons presented in Fig. \ref{gross},  we can state  that the 
DIT/GEMINI calculation is  able to provide  a satisfactory quantitative description
of the observed gross  distributions.
Also, as we will see below, it does a fair job in predicting the absolute 
values of the production cross sections (except for the very n-rich isotopes,
as will be pointed out).

In Fig. \ref{z1}, the  mass distributions for elements Z=35--30 are presented. 
In this figure, the full circles are the present data,  corrected 
for the spectrometer  acceptance as discussed earlier.
The open squares are  the predictions of the DIT/GEMINI 
calculation. The dotted  lines are  the predictions of the EPAX 
parametrization \cite{EPAX} of relativistic fragmentation cross sections 
and are  plotted here  for comparison. Note that, in high-energy fragmentation, 
nucleon-pickup products are not produced  or, at best, are highly  suppressed
compared to lower energy peripheral collisions \cite{Weber}.
As we see, neutron-rich nuclides are produced with large cross  sections. 
Apart from proton-removal products, neutron pick-up fragments are produced.
For example, for the case of selenium (Z=34), which corresponds to the removal of 2 protons
from the projectile, up to 4 neutrons are seen to be picked-up to produce  $^{88}$Se
(with cross section of 0.3 $\mu$b). Similarly, for germanium (four-proton-removal from the 
projectile), up to two neutrons are picked-up to give $^{84}$Ge (cross section 0.4 $\mu$b).

\begin{figure}[p]                            

\includegraphics[width=10.1cm,height=15.5cm]{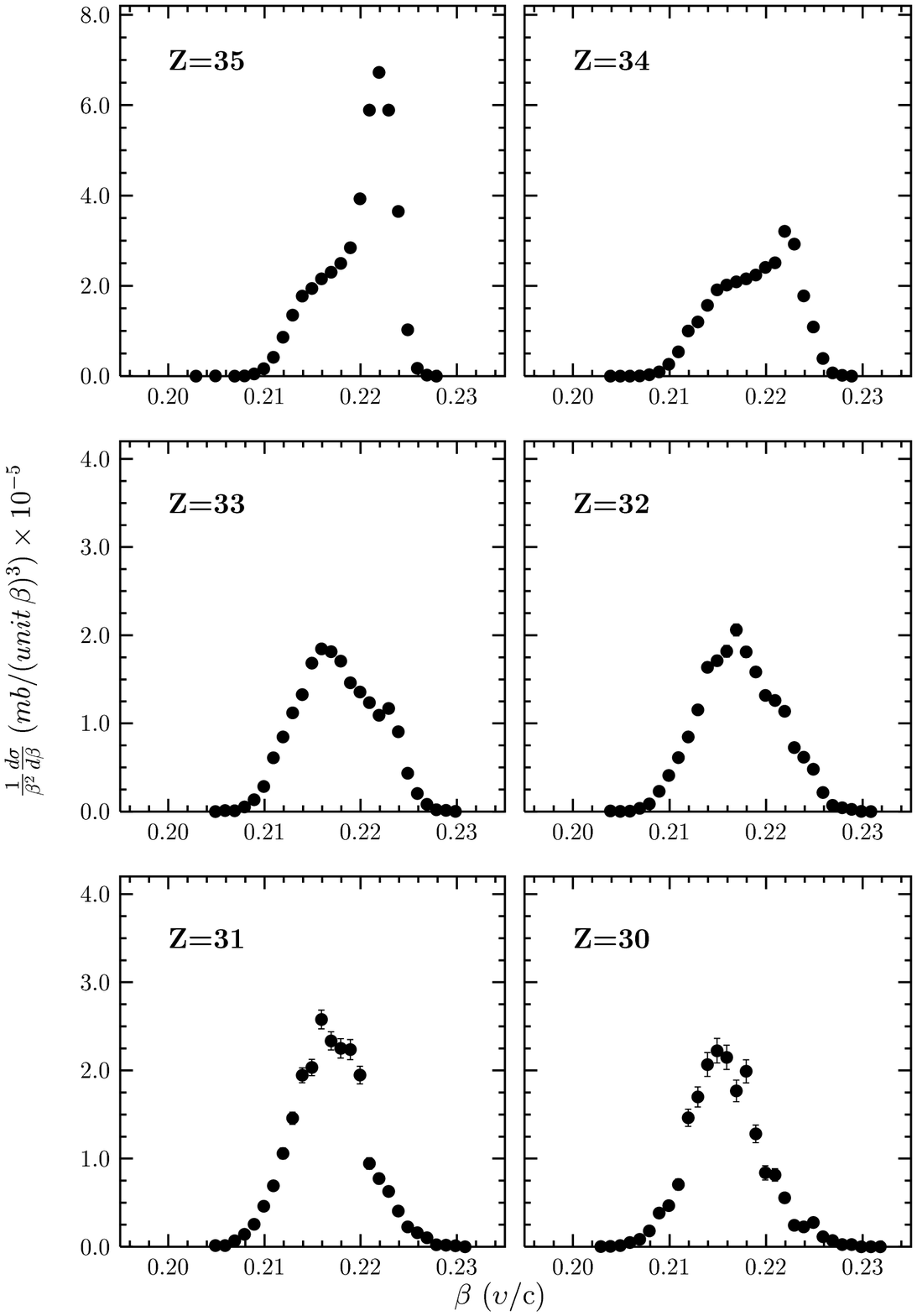}
\centering
\vspace{0.0cm}

\caption{ Velocity distributions for elements Z=35--30
          (see text). }
\label{vel}
\end{figure}

For near-projectile elements, an enhanced  production
of neutron rich isotopes is observed relative to the expectations of the 
EPAX parametrization. Also, the DIT/GEMINI calculation, while able to describe
rather well the n-deficient and the central part of the distributions, 
fails to describe the n-rich sides of the distributions for elements
above zinc (Z=30).
Since  this  enhancement is observed for near projectile elements,
it should be a result of very peripheral collisions, where the nucleon exchange
is restricted near the surface of the projectile and the target.
To further examine this statement, we present in Fig. \ref{vel} the velocity 
distributions for elements Z=35--30. These distributions have two components: 
a narrow one peaking close to the beam velocity (quasielastic) and another 
wider component peaking at lower velocity (deep inelastic component).
The quasielastic  component progressively decreases for fragments with lower Z.
As expected, the production of the most neutron-rich isotopes is associated with 
the quasielastic component (also verified in  two-dimensional correlations of 
A with velocity). Using velocity vs impact parameter correlation from the DIT/GEMINI 
calculation, we found that these peripheral  events correspond to a projectile--target  
overlap not exceeding 1--1.5 fm.

Qualitatively, the enhanced production of neutron rich fragments from  
$^{86}$Kr may be understood  by considering the peripheral character of the collisions
and the particular structure of the projectile and the target. The projectile has a 
compact neutron distribution (N=50 closed shell)
and, reversely,  the target has a compact proton distribution (Z=28 closed shell)
and a more diffuse neutron distribution  
(N=36, partially filled  neutron shell with 8 neutrons above the closed shell N=28).
In peripheral collisions, an overlap of the  $^{86}$Kr projectile surface 
with the neutron-rich $^{64}$Ni surface (neutron ``skin'') will lead to a local 
nucleon redistribution and N/Z equilibration, favoring the production of 
more neutron-rich fragments near the projectile.
For larger projectile--target overlaps this effect washes out and the 
observed fragments have more or less cross sections as described 
by DIT/GEMINI. Interestingly,  they are in general agreement with EPAX,
despite  the vastly different primary ineraction mechanism, indicating 
that such products result from long evaporation chains of highly excited 
primary nuclei. It should be noted that in the DIT code, the  nuclei are assumed spherical
with homogeneous proton and neutron density distributions.  Incorporating  realistic proton 
and neutron density distributions  in the DIT code may improve the  capability to 
describe peripheral collisions  in  projectile--target combinations where the neutron--proton 
profile of nuclear surface can play a role,  as in the present case of the 
$^{86}$Kr+$^{64}$Ni system.

From a practical standpoint,   the large production  cross section
of neutron-rich nuclides, and more importantly the neutron pick-up
possibility can render these reactions a useful route to produce
extemely neutron-rich nuclides.
It may be noted that neutron pick-up cross sections are also large in 
the case of multinucleon transfer reactions close to the Coulomb barrier
\cite{Corradi}. However, these reactions have very wide angular, velocity
and ionic charge distributions,  in addition to the necessity  of rather 
thin targets (around 1 mg/cm$^2$).
Around the Fermi energy, however, inverse kinematic reactions have 
angular and ionic charge state distributions that can be efficiently dealt 
with using a large acceptance  spectrometer 
(e.g. MARS in the present study). Also the energies are high enough to 
allow moderately thick targets (10--30 mg/cm$^2)$.

Using the present cross sections,  we can make estimates of rare isotope production 
rates from  intense beams at this energy regime. 
As examples, we present in Table I the cross sections and production rates
for the most n-rich nuclides of Se (Z=34) and Ge (Z=32). 
The experimental cross sections of this work are given in the second column, while the  
the third  column gives the cross sections predicted by the DIT/GEMINI 
calculation. The fourth column gives the cross sections  measured in the reaction
$^{86}$Kr(500 MeV/nucleon) + Be \cite{Weber}. We observe, as already discussed, the inability of 
the present simulation to reproduce the measured cross sections of the most n-rich 
nuclides. Also, we note  that in relativistic peripheral collisions, up to one neutron can be 
picked up from the target with very low cross section.
The rates given in the last column of the table  were calculated using the 
measured cross sections and  
assuming a beam of 100 pnA  $^{86}$Kr at 25 MeV/nucleon  striking 
a  20 mg/cm$^{2}$ $^{64}$Ni target.
Such yields of very neutron rich isotopes may enable a variety of nuclear structure 
and nuclear reaction  studies in the Fermi energy regime. 
These rare  isotopes may be separated  in flight with a large acceptance  separator
or can be stopped  and collected in a gas cell with the possibility of  subsequent charge 
breeding and reacceleration \cite{TAMU}.

Finally, another interesting possibility is the use of such reactions 
as a second stage in two-stage rare isotope production schemes.
For example, a beam of $^{90}$Kr  from an ISOL facility can be accelerated around the 
Fermi energy and subsequently strike a $^{64}$Ni target to produce 
very n-rich nuclides that may be separated and studied in flight.
To estimate the rates of such reaction products the present cross sections can be 
used as a first approximation. However, a quantitative prediction will be possible 
after improving our  present description of these peripheral collisions 
between n-rich nuclei.
    
 \section{Summary and conclusions}

In the present study,  the yields and velocity distributions of projectile-like 
fragments  from the reaction of 25 MeV/nucleon $^{86}$Kr + $^{64}$Ni 
were  measured using the MARS recoil separator at Texas A\&M,  with 
special focus on the neutron rich isotopes. Proton-removal and 
neutron pick-up isotopes have been observed with substantial cross sections. 
A model of deep-inelastic transfer (DIT) for the primary interaction stage  
and the  statistical evaporation code GEMINI for the deexcitation stage
have been used to describe the properties of the product  distributions.
The results have also been  compared with the high-energy fragmentation 
parametrization EPAX. 
The experimental data show an enhancement  in the production of  n-rich 
isotopes close to the projectile relative to the predictions of DIT/GEMINI 
and the expectations of EPAX. We  attributed this enhancement  to the 
effect of the target $^{64}$Ni neutron ``skin'' in  peripheral interactions of 
$^{86}$Kr with $^{64}$Ni.
The large cross sections of such reactions near   the Fermi energy, 
involving  peripheral nucleon exchange between the projectile 
and the target, suggest that not only the N/Z of the projectile and the
target, but also the N/Z distribution at the surface (i.e. neutron ``skin'') 
may be properly  exploited in the production of rare neutron-rich isotopes.
This synthesis approach  may  provide  a fruitful pathway to  extremely 
neutron rich  nuclei towards  the neutron-drip line.


\section{Acknowledgement}


We would like to thank A. Sanzhur for useful discussions  regarding
neutron density distributions and relevant calculations. We gratefully 
acknowledge the support of the  operations staff of the Cyclotron Institute
during the measurements.
Financial support for this work was given, in part, by the U.S. Department 
of Energy under Grant No. DE-FG03-93ER40773 and by the Robert A. Welch 
Foundation under Grant No. A-1266.





\newpage

%
%

\begin{table}[t]                     
\caption{ Cross sections and rates (last column) of Se and Ge isotopes
          from $^{86}$Kr fragmentation. 
          For the rates,  the cross section data of this work are used and 
          a primary beam of $^{86}$Kr (25 MeV/nucleon) of intensity 100 pnA  is 
          assumed to  interact with  a $^{64}$Ni target  of 20 mg/cm$^{2}$ thickness 
          (see text).
                   }
\vspace{0.5cm}
\begin{center}
\begin{tabular}{cccccc}
\hline
\hline
 Rare      &        &  Cross Sections &                 &                    &  Rate (sec$^{-1}$)  \\    
 Isotope:  &Reaction&  Experiment:    & Calculated      & High Energy        &        \\
           &channel &  This work      & DIT/GEMINI      & Data \cite{Weber}  &        \\  \hline
           &        &                 &                 &                    &        \\  
    
$^{84}$Se  & -2p+0n &   4.7 mb        &  0.8 mb         &     2 mb           & 5.6$\times$10$^{5}$  \\

$^{85}$Se  & -2p+1n &   900 $\mu$b    &   60 $\mu$b     &     8 $\mu$b       & 1.1$\times$10$^{5}$  \\

$^{86}$Se  & -2p+2n &   100 $\mu$b    &    3 $\mu$b     &     --             & 1.2$\times$10$^{4}$  \\

$^{87}$Se  & -2p+3n &    14 $\mu$b    &   --            &     --             & 1.7$\times$10$^{3}$  \\

$^{88}$Se  & -2p+4n &   0.3 $\mu$b    &   --            &     --             &   3$\times$10$^{1}$  \\

$^{82}$Ge  & -4p+0n &   22  $\mu$b    &    2  $\mu$b    &     3    $\mu$b    & 2.6$\times$10$^{3}$  \\

$^{83}$Ge  & -4p+1n &  2.3  $\mu$b    &    --           &     0.04 $\mu$b    & 2.8$\times$10$^{2}$  \\

$^{84}$Ge  & -4p+2n &  0.4  $\mu$b    &    --           &     --             & 4.8$\times$10$^{1}$  \\

           &        &                 &                 &                    &                      \\
\hline
\hline
\end{tabular}
\end{center}
\end{table}
%
%
%
%

\begin{thebibliography}{00}

\bibitem{RIA} Long-Range Plan for Nuclear Physics (1996),  DOE--NSF Nucl. Sci. Advisory 
              Committee, accessible at  www.er.doe.gov/production/henp/np/nsac/lrp.html.

\bibitem{EURISOL} Radioactive Nuclear Beam Facilities, NuPECC Report, April 2000;
                  see also EURISOL web page: www.ganil.fr/eurisol/


\bibitem{Geissel} H. Geissel and G. Munzenberg, Annu. Rev. Nucl. Part. Sci. 45 (1995) 163.
\bibitem{Morrissey}   D.J. Morrissey and B. M. Sherrill, 
                      Phil. Trans. R. Soc. Lond. A  356 (1998) 1985.
 \bibitem{Friedman}   W.A. Friedman, M.B. Tsang, D. Bazin, and W.G. Lynch, 
                      Phys. Rev.  C62 (2000) 064609-1.
\bibitem{EPAX}    K. Summerer and B. Blank,  Phys. Rev.  C61 (2000) 034607.
\bibitem{Weber} M. Weber et al., Nucl. Phys. A578 (1994) 659.
%
\bibitem{Corradi} L. Corradi et al.,  Phys. Rev.  C 59 (1999) 261.
\bibitem{Tarasov} O.B. Tarasov et al., Nucl. Phys. A629 (1998) 605.
\bibitem{Artukh}  A.G. Artukh et al., Nucl. Phys. A 176 (1971) 284.
\bibitem{Volkov}  V.V. Volkov,  Phys. Rep. 44 (1978) 93.
\bibitem{IYLee}   I.Y. Lee et al.,  Phys. Rev. C 56  (1997) 753.
\bibitem{Bacri} Ch.O. Bacri et al., Nucl. Phys. A 555 (1993) 477.
\bibitem{Bazin}
  D. Bazin et al.,  Nucl. Phys. A 515 (1990) 349.
\bibitem{Pfaff} R. Pfaff et al.,  Phys. Rev. C51 (1995) 1348.
\bibitem{MARS}   R.E. Tribble, R.H. Burch and C.A. Gagliardi,
                 Nucl. Instr. and  Meth. A 285 (1989) 441.
\bibitem{Wilcke} W.W. Wilcke et al.,
                 At. Data Nucl. Data Tables {\bf 25}, 389 (1980).

\bibitem{Greg}  G. Chubarian, private communication.
            
\bibitem{Hubert}
F. Hubert, R. Bimbot and H. Gauvin, Atom. Data and Nucl.
Data Tables  46 1 (1990) and Nucl. Instrum. Methods  B36 (1989) 357.
\bibitem{Leon} Leon et al.,  At. Data Nucl. Data Tables 69  (1998) 217.

\bibitem{AuPLF}
 G.A. Souliotis et al.,  Phys. Rev.  C57  (1998) 3129.

\bibitem{DIT}
 L. Tassan-Got, and C. Stefan,  Nucl. Phys. A524 (1991) 121.

 \bibitem{MV0}
  M. Veselsky et al.,  Phys. Rev.  C62  (2000) 064613.
\bibitem{GEMINI}
 R. Charity, et al., Nucl. Phys.  A483  (1988) 391.
 The version of GEMINI included modifications made up to July, 1998.
\bibitem{Moretto}
 L.G. Moretto, Nucl. Phys. A247 (1975) 211.
\bibitem{Lestone}
 J. Lestone, Phys. Rev.  C52  (1995) 118.
\bibitem{Marmier}
 P. Marmier and E. Sheldon, Physics of Nuclei and Particles, Volume I (Academic,
 New York, 1970) p. 15.
\bibitem{TAMU}
 ``A Proposed Facility Upgrade for the  Texas A\&M Univ. Cycloton Institute'',
 accessible at:  http://cyclotron.tamu.edu/facility$_{-}$upgrade.htm


\end{thebibliography}
\end{document}